\documentclass[11pt]{article}
\usepackage{xspace}
\usepackage{graphicx}
\usepackage{amsmath}
\usepackage{amssymb}
\usepackage{color}

\definecolor{Red}{rgb}{1,0,0}
\definecolor{Green}{rgb}{0,1,0}
\definecolor{Blue}{rgb}{0,0,1}
\definecolor{Black}{rgb}{0,0,0}

\def\beq{\begin{equation}}
\def\eeq#1{\label{#1}\end{equation}}
\def\eeqn{\end{equation}}

\def\beqa{\begin{eqnarray}}
\def\eeqa#1{\label{#1}\end{eqnarray}}
\def\eeqan{\end{eqnarray}}

\let\bar=\overbar

\def\Dslash{\not{\hbox{\kern-4pt $D$}}}
\def\dslash{\not{\hbox{\kern-2pt $\del$}}}

\def\msb{{\bar{\ssstyle M \kern -1pt S}}}

\def\Title#1{\begin{center} {\Large {\bf #1} } \end{center}}

\begin{document}

\Title{Searching for Sterile Neutrinos at MINOS}

\bigskip\bigskip


\begin{raggedright}  

{\it Ashley Timmons\index{Timmons, A.},\\
School of Physics and Astronomy\\
The University of Manchester\\
M13 9PL Manchester, UK}\\

\end{raggedright}
\vspace{1.cm}


\section{Introduction}

\begin{figure}[!ht]
\begin{center}
\includegraphics[trim=0cm 1.0cm 0.0cm 1cm, scale=0.45]{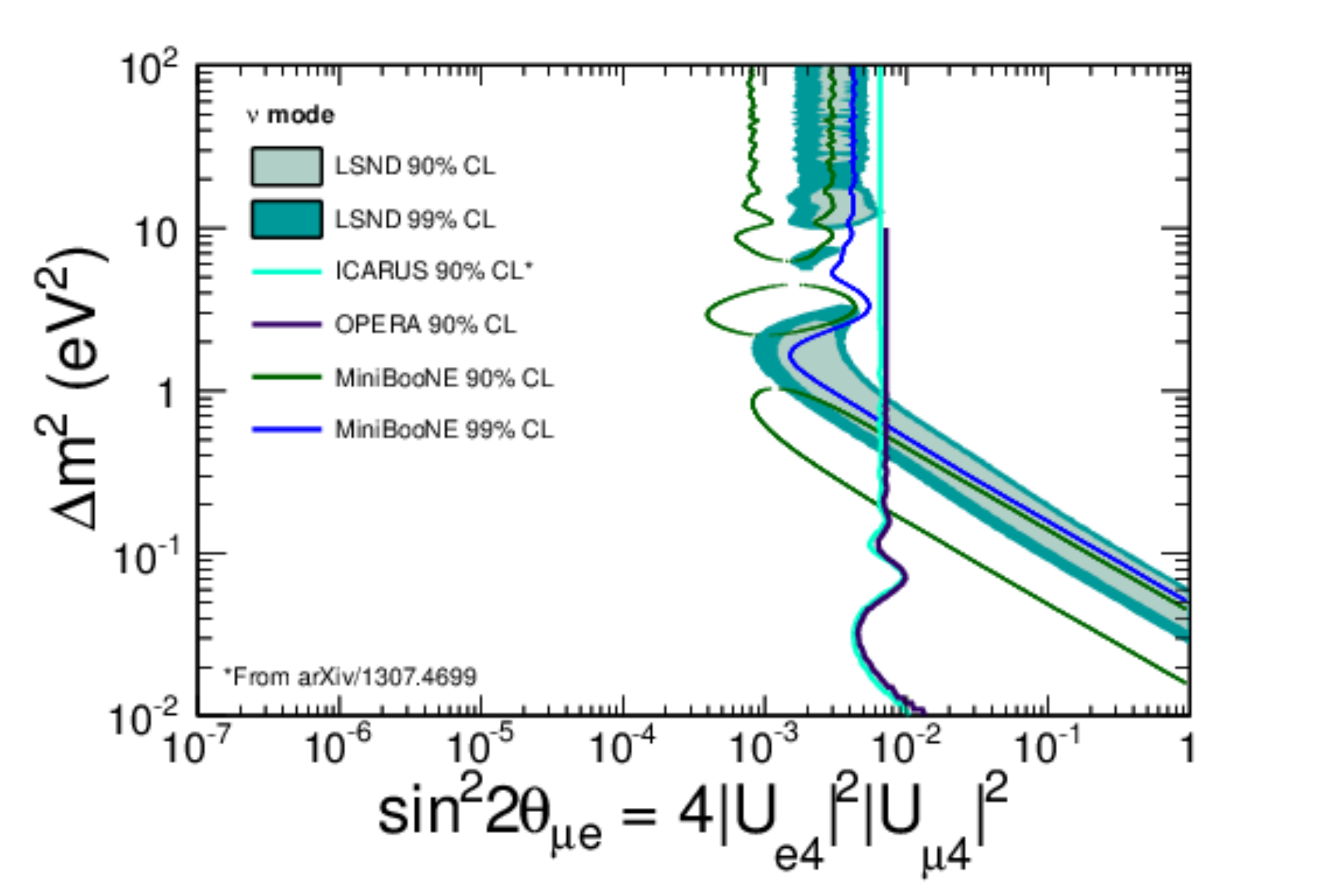}
\caption{Showing the tension between various confidence regions for experiments searching for a sterile signal.}
\label{fig:example}
\end{center}
\end{figure}

There have been several anomalous results within the neutrino community that have questioned our understanding of neutrinos. The Liquid Scintillator Neutrino Detector (LSND) and MiniBooNE short-baseline experiments saw an excess of electron antineutrinos that can not be explained using the current three-flavour model \cite{LSND, MINI}.  One explanation is the addition of one or more neutrino types which would oscillate amongst the three active neutrino flavours. Figure \ref{fig:example} shows the allowed regions for a third mass splitting $\Delta m^{2}$ (the difference between neutrino states) and additional mixing angle to accommodate a fourth neutrino. However, measurements of the decay width of the $Z$ boson show that only three light active neutrino flavours exist \cite{LEP}. Any additional neutrinos must be sterile, not interacting via the weak interaction.

By looking for perturbations from the three-flavour oscillation formalism in $\nu_{\mu}$ charged current (CC) events and a deficit in neutral current events (NC), MINOS can search for sterile neutrinos in the parameter space suggested by LSND and MiniBooNE. This MINOS analysis uses a 3+1 model to quantify the consistency with the three flavour neutrino model in data from a $\nu_{\mu}$ beam and a baseline of 735 km. This complements other previous experimental searches for sterile neutrinos in the $\nu_{\mu} \rightarrow \nu_{e}$ appearance channel.

\section{The MINOS Experiment}
MINOS \cite{MINOSNIM} is a two-detector on-axis long-baseline neutrino experiment based at Fermilab. The experiment samples the NuMI energy beam, for this analysis the data was taken with the NuMI beam configured to produce predominantly muon neutrinos, produced by 120 GeV protons incident on a segmented graphite target. The Near Detector (ND) measures the neutrino energy spectrum and is located 1 km from the production target at Fermilab. The Far Detector (FD), again measures the energy spectrum 734 km further away at Soudan Underground Laboratory, in northern Minnesota. Both detectors are tracking, sampling calorimeters consisting of alternate planes of steel and scintillator strips.  Constructed with a similar design to allows uncertainties affecting both detectors in similar ways to cancel out to first order, such as cross section and flux uncertainties.

\section{Sterile Neutrino Analysis}
A 3+1 model gives rise to an additional mass splitting $\Delta m^{2}_{43}$ along with three new mixing angles $\theta_{14}, \theta_{24}$ and $\theta_{34}$. A traditional neutrino oscillation analysis assumes that the ND provides a measurement of the neutrino energy spectrum before oscillations. However, for mass splittings above $\sim 1 \,\, \text{eV}^{2}$ oscillations cause disappearance of muon neutrinos in the ND. By using the F/N ratio for the fit, we are able to account for this and extend our sensitivity over five orders of magnitude $\Delta m^2_{43}$.

\begin{figure}[!ht]
\begin{center}
\includegraphics[scale = 0.37]{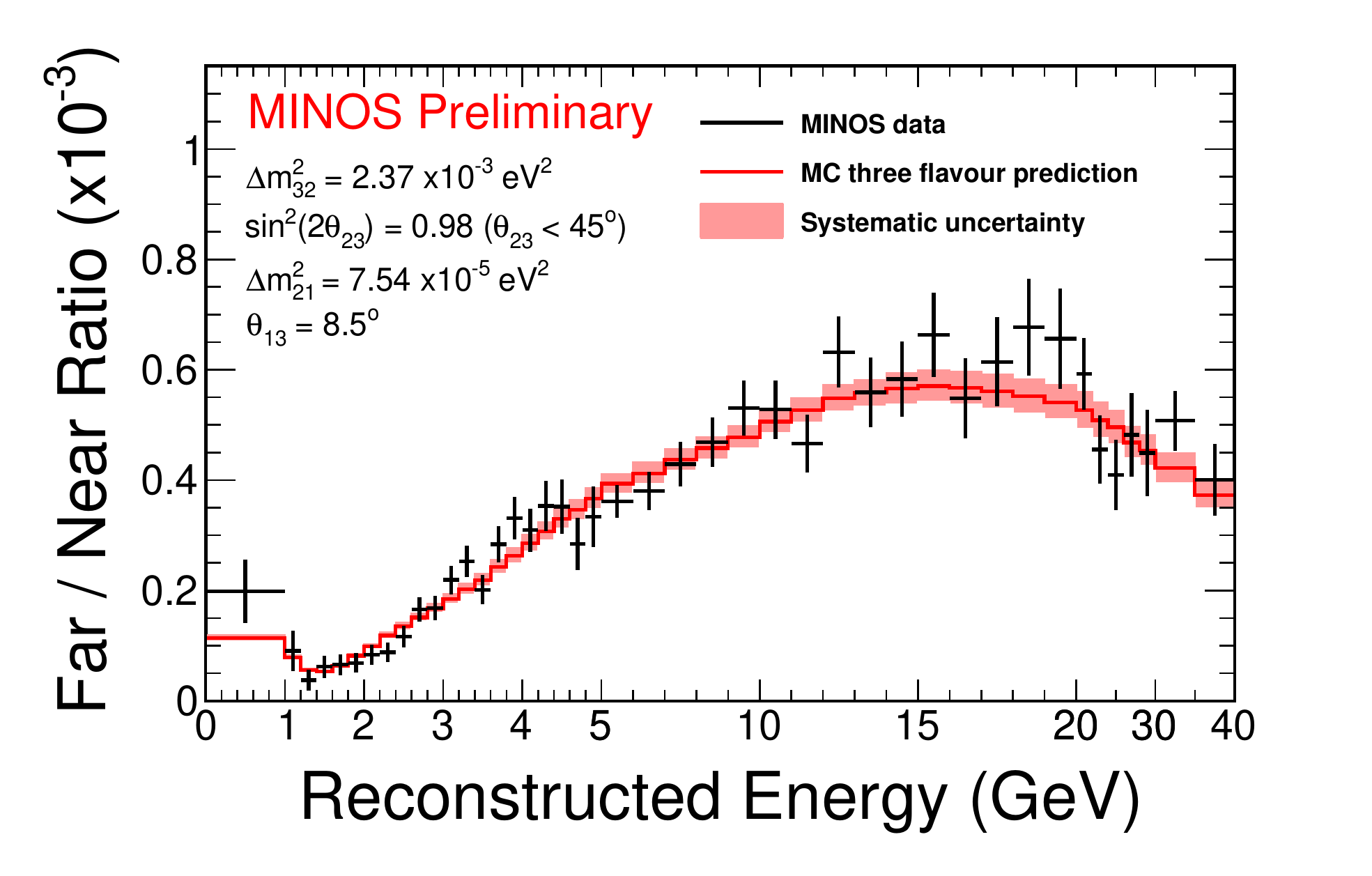}
\includegraphics[scale = 0.37]{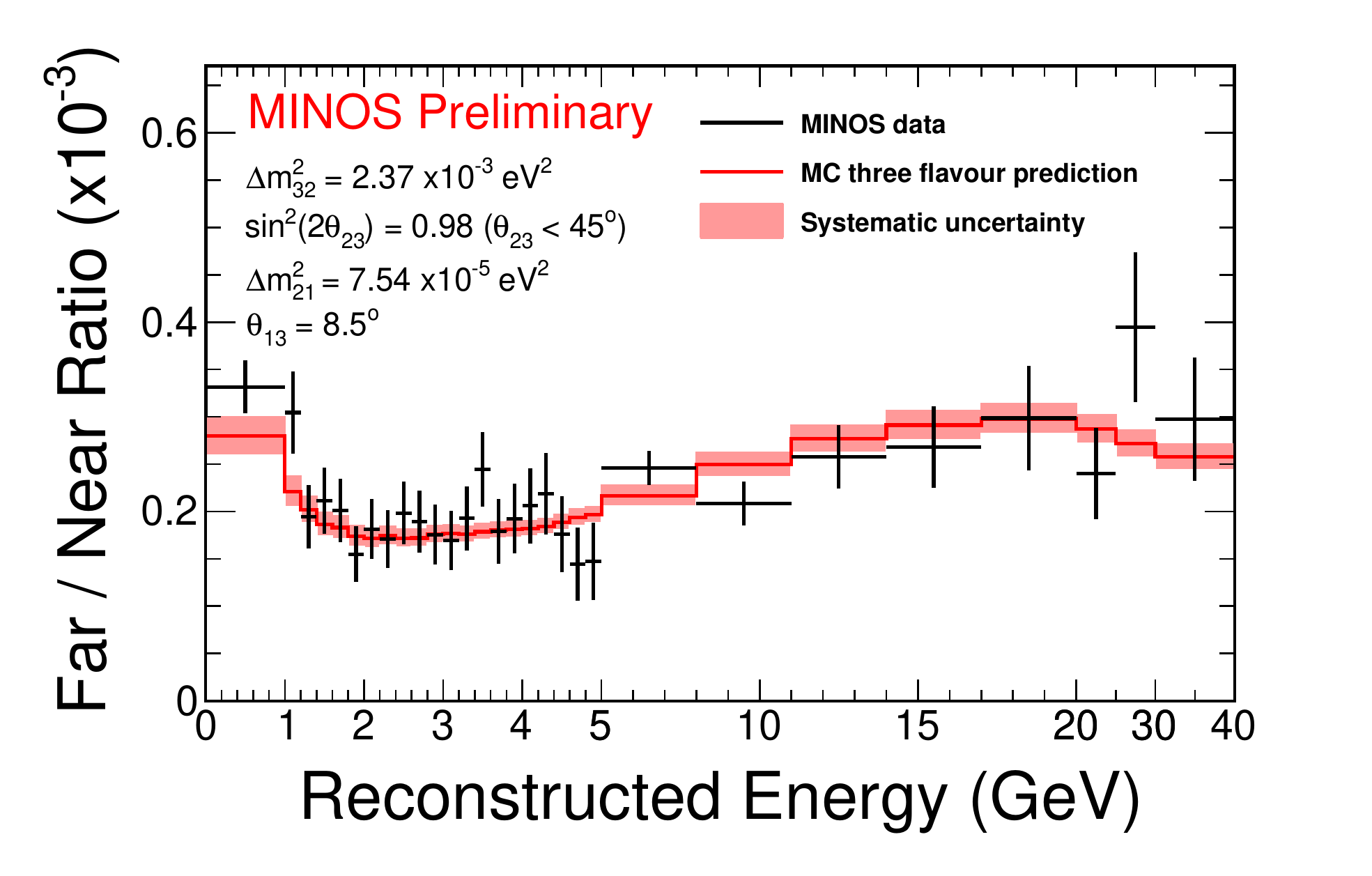}
\caption{Data and Monte Carlo Far/Near ratios for CC (left) and NC (right) selected events.}
\label{fig:Ratios}
\end{center}
\end{figure}

This analysis includes the systematic uncertainties on the predicted F/N ratio for each bin of energy in the form of a covariance matrix. MINOS currently fits for three parameters $\Delta m^2_{32}, \theta_{23} \,\, \text{and} \,\,  \theta_{34}$ while searching over the parameter space $\Delta m^{2}_{43} - \sin^{2}(2\theta_{24})$. This analysis fixes the values of $\Delta m^2_{21}, \theta_{12}, \theta_{13} \,\,\text{and}\,\, \theta_{14}$ to the global best fit due to having a negligible impact on the fit. A $\chi^{2}$ is calculated from both the CC and NC F/N ratios along with a penalty term for the total number of events at the ND as seen in equation \ref{chisq}.

\begin{align}
\label{chisq}
\chi^{2} = \mathbf{V}^{T} \mathbf{C}^{-1} \mathbf{V} \,\,\,+ \frac{(N_{D}-N_{MC})^{2}}{\sigma_{MC}^{2}}, 
\end{align}

\noindent where the vector $\mathbf{V}$ is the difference between the data $D$ and prediction $MC$ for the $i^{\text{th}}$ bin of reconstructed energy $V^{T}_{i} = (D_{i} - MC_{i})$. The inverse of the covariance matrix is given by $\mathbf{C}^{-1}$.

This analysis places constraints over 4 orders of magnitude in the sterile mass splitting $\Delta m^2_{43}$, ranging between $10^{-3} - {10^{2}} \,\, \text{eV}^{2}$. The distance from the neutrino vertex in the ND to the meson decay point is taken into account when computing the oscillations; when sterile neutrinos start to oscillate at the ND this becomes an effect due to the 675 m long decay pipe. The data surface obtained was corrected using the \mbox{Feldman-Cousins} unified approach \cite{FC}. 

\begin{figure}[!ht]
\begin{center}
\includegraphics[trim=2cm 0.5cm 1.0cm 1cm, scale=0.55]{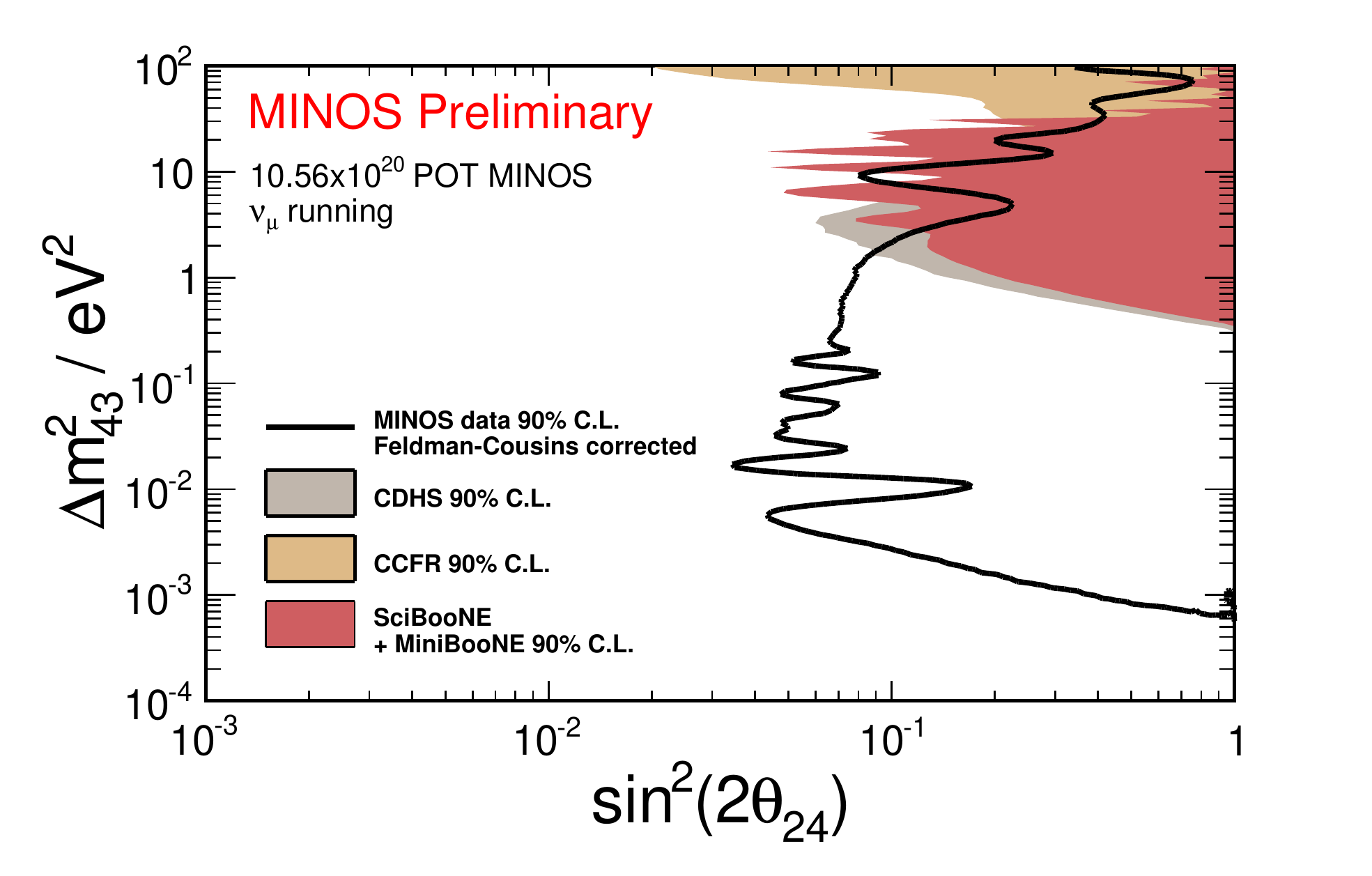}
\caption{Showing the Feldman-Cousins corrected 90\% C.L. disappearance exclusion obtained on the $\theta_{24}$ sterile mixing angle. Everything to the right of the black line is excluded.}
\label{fig:DataLimits}
\end{center}
\end{figure}

\begin{figure}[!ht]
\begin{center}
\includegraphics[trim=6cm 0cm 0cm 8cm, scale=0.2]{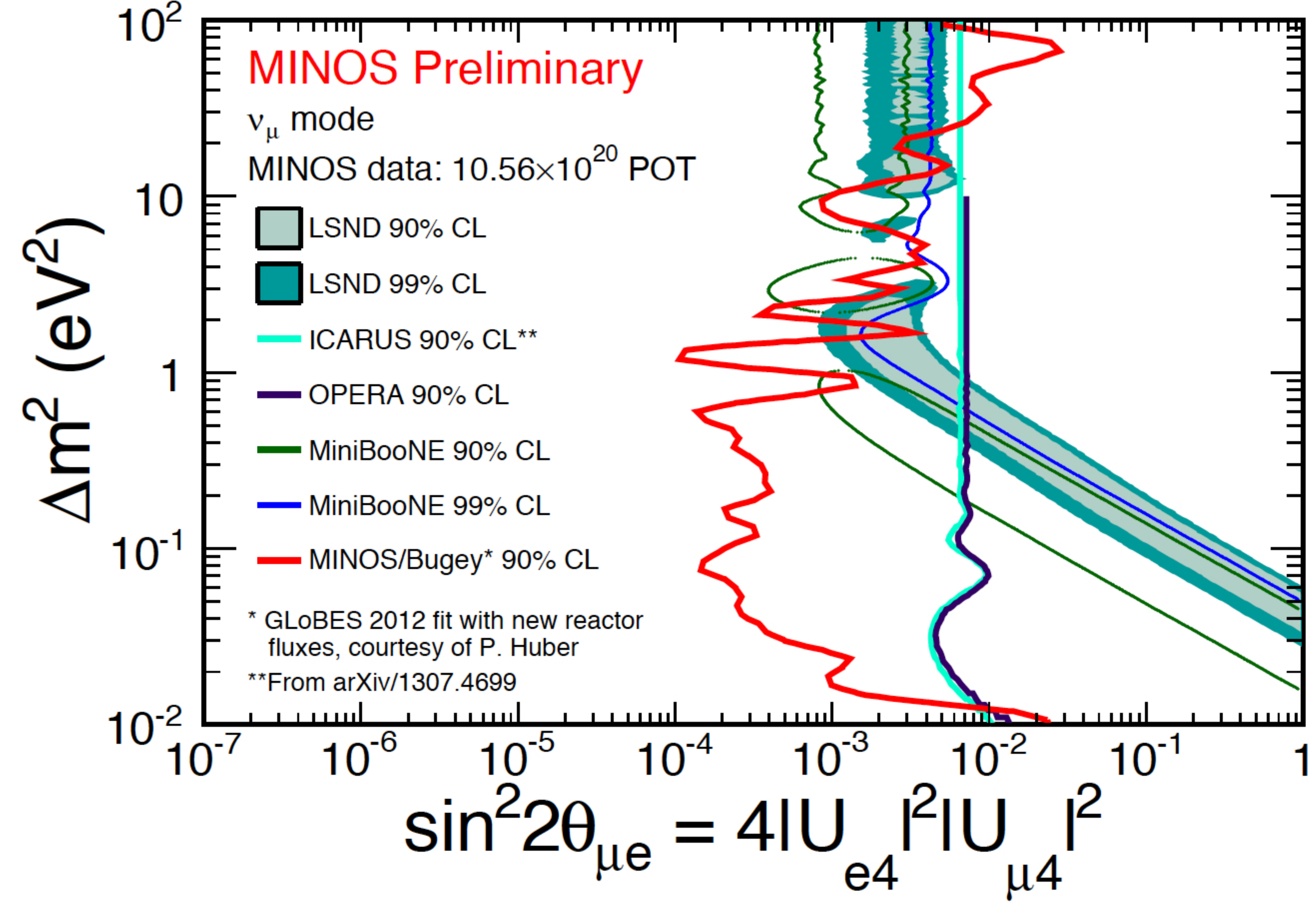}
\caption{This plot shows the results of combining the MINOS limit with the Bugey 90\% C.L. limit on $\nu_{e}$ disappearance \cite{bugey}}
\label{fig:budgyLimits}
\end{center}
\end{figure}

\section{Summary}
By performing a fit on the ratio between the FD and ND energy spectra for both CC and NC events, MINOS can search for a sterile signature in the region of parameter space suggested by LSND and MiniBooNE. MINOS assumes a 3+1 model and sets a limit spanning five orders of magnitude in $\Delta m^2_{43}$ and places one of the strongest constraints on $\Delta m^2_{43} \lesssim 1 \,\, \text{eV}^{2}$. Systematics are handled by creating a covariance matrix containing the uncertainties which would affect the F/N reconstructed energy spectrum and a $\chi^{2}$ fit was performed to the data. This result adds to the tension of null results from sterile disappearance searches compared to signals discovered in the appearance channel as seen in figure \ref{fig:budgyLimits}.

\bigskip
\section{Acknowledgments}
This work was supported by the U.S. DOE; the U.K. STFC; the U.S. NSF; the State and University of Minnesota; the University of Athens, Greece; and Brazil's FAPESP, CAPES, and CNPq. We thank the staff of Fermilab for their invaluable contributions to the research of this work.

\end{document}